\documentclass[aps,twocolumn,floats,superscriptaddress,showpacs]{revtex4}
\usepackage{amsmath,bm}%
\usepackage{graphicx}%

\begin{document}
\title{Cluster Emission and Phase Transition Behavior  in the Nuclear Disassembly}
\author{Y. G. Ma}
\thanks{Present address: Cyclotron Institute, Texas A $\&$ M University, College Station,
 TX 77843-3366, USA. Email: ygma@cyclotronmail.tamu.edu. http://ma.tamu.edu}
\affiliation{Shanghai Institute of Nuclear Research, Chinese
Academy of Sciences, P.O. Box 800-204, 201800 Shanghai, CHINA}
%\date{\today}
\date{Oct. 1999}
\begin{abstract}
The features of the emissions  of light particles ($LP$), charged
particles ($CP$), intermediate mass fragments ($IMF$) and the
largest fragment ($MAX$)  are investigated for $^{129}Xe$ as
functions of temperature and "freeze-out" density in the
frameworks  of isospin dependent lattice gas model and classical
molecular dynamics model. Definite turning points for the slopes
of average multiplicity of $LP$, $CP$ and $IMF$, and of the mean
mass of the largest fragment ($A_{max}$) are shown  around a
liquid gas phase transition temperature  and while  the largest
variances of the  distributions  of $LP$, $CP$, $IMF$, and $MAX$
appear there. It indicates that the cluster emission rate can be
taken as a probe of nuclear liquid gas phase transition and while
the largest fluctuation is simultaneously accompanied at the point
of the phase transition within lattice gas model and molecular
dynamics model as the traditional thermodynamical  theory requires
when a phase transition occurs.
\end{abstract}

%\vskip2pc]

\maketitle

%\newpage
\section{Introduction}

Phase transition and critical phenomenon is an extensive debating
subject in  nature sciences. Recently, the same concept was introduced
into the microscopic systems, such as  in atomic cluster \cite{Bert?} and
nuclei \cite{Finn82}. The break-up of nuclei due to violent collisions
into several intermediate mass fragments, can be viewed as critical
phenomenon as observed in fluid, atomic, and other systems.
It prompts the possible signature on the liquid gas phase transition in the
nuclear system. One fact is that the sudden opening of the
multifragmentation \cite{Biza93} and vaporization \cite{Rive96} channels
can be interpreted as the signature of the boundaries of phase mixture
\cite{Gros90}. In addition, the fact that
the nuclear caloric curve \cite{Poch95,Haug96,Mayg97,Serf98,Tsang97,Xi97}
in a certain excitation energy range shows a plateau gives a possible
indication of a first order phase transition \cite{Poch95} as predicted in the
framework of statistical equilibrium models \cite{Bond98}.
On the other hand, the extraction of
critical exponents  in the charge or mass  distribution of the
multifragmentation system \cite{Gilk94} can be explained as an
evidence of the phase transition.

Several theoretical models have been developed to treat such a
phase transition in nuclear disassembly. These models are
percolation model \cite{Camp88,Baue88}, lattice gas  model (LGM)
and statistical multifragmentation model \cite{Bond95} etc. In
this work, what we are interested in is the lattice gas model. It
is a simple short range interaction model \cite{Yang52}, but it
can  be successfully applied to nuclear systems with isospin
symmetry and asymmetry
\cite{Jpan95,Jpan96,Camp97,Carmona,Gupta97,Sray97,Biro86,Sama,Mull97,Gulm98,Ma99}.
LGM is carried assuming the system is in a  "freeze-out" density
$\rho_f$ with thermal equilibrium at temperature $T$. Previous
calculations \cite{Gupta97} with LGM showed that there exists a
phase transition for the finite nuclear systems by studying the
effective power law parameter ($\tau$)  of cluster  mass or charge
distribution \cite{Fish67,Lit93}, their second moments ($S_2$)
\cite{Camp88}, and the specific heat. More recently, we proposed
two novel criteria, namely multiplicity information entropy ($H$)
and nuclear Zipf's law to diagnose the onset of liquid gas phase
transition in the framework of the isospin dependent LGM (I-LGM)
and isospin dependent classical molecular dynamics (I-CMD) model
\cite{Maprl}.

In this work, we will show that the emission rate of clusters is a
useful tool to diagnose the nuclear liquid gas phase transition,
and while the largest fluctuation of cluster multiplicities is
simultaneously revealed at the point of  the phase transition by
investigating  the features of light particles, charged particles,
intermediate mass fragments and the largest fragment of
disassembling source in the frameworks of I-LGM and I-CMD.

The paper is organized as following. Sec.II gives the descriptions
of I-LGM and I-CMD. Results and discussions are shown in Sec. III
where the multiplicities of cluster emissions, their slopes and
their fluctuations are investigated. The influence of the
"freeze-out" density on cluster emission is also presented in the
framework of I-LGM, and comparisons of the results between I-LGM
and I-CMD at a given "freeze-out" density is followed. Finally the
conclusion is given in Sec.IV.

\section{Description of Models}
\subsection{Isospin dependent lattice gas model}

The lattice gas model of Lee and Yang \cite{Yang52}, where the
grand canonical partition function of a gas with one type of atoms
is mapped into the canonical ensemble of an Ising model for  spin
1/2 particles, has successfully described the liquid-gas phase
transition for atomic system. The same model has already been
applied to microscopic nuclear system. Detailed descriptions for
the features of LGM can be found in some literatures, eg. papers
of Pan and Das Gupta et al. \cite{Jpan95,Jpan96},  of Campi and
Krivine \cite{Camp97}, of Carmona, Richert and Tarancon
\cite{Carmona} and of Chomaz and Gulminelli \cite{Gulm98} etc.
Here we will adopt the model developed by Pan and Das Gupta
\cite{Jpan95,Jpan96}. In order to better understand the context of
this work, the models are  described below.

In the LGM, $A$ (=$N + Z$) nucleons with an occupation number $s$
(i.e. "spin") which is defined as $s$ = 1 (-1) for a proton (neutron)
or $s$ = 0 for a vacancy, are placed in the $L$ sites of 3 dimension
lattice. Nucleons in the nearest neighboring sites interact
with an energy $\epsilon_{s_i s_j}$. The hamiltonian of the system
is written as
\begin{equation}
 H = \sum_{i=1}^{A} \frac{P_i^2}{2m} -
\sum_{i < j} \epsilon_{s_i s_j}s_i s_j .
\end{equation}
The interaction constant $\epsilon_{s_i s_j}$ is related  to the binding
energy of the nuclei.
In order to incorporate the isospin effect in the lattice gas model,
the short-range interaction constant $\epsilon_{s_i s_j}$
is chosen to be different between the nearest neighboring like nucleons
and unlike nucleons:
\begin{eqnarray}
 \epsilon_{nn} \ &=&\ \epsilon_{pp} \ = \ 0. MeV \nonumber , \\
 \epsilon_{pn} \ &=&\ - 5.33 MeV,
\end{eqnarray}
 which indicates the repulsion between the nearest-neighboring like
 nucleons and the attraction between the nearest-neighboring unlike
 nucleons. This kind of isospin dependent interaction incorporates,
 to a certain extent, Pauli exclusion principle and effectively
 avoids producing unreasonable clusters, such as di-proton and
di-neutron clusters, etc.  The disassembly of the system is to be
calculated at an assumed "freeze-out" density $\rho_f$ =
$(A/L)\rho_0$, where
$\rho_0$ is the normal nucleon density, beyond $\rho_f$ nucleons
are too far apart to interact.

In the model, $N + Z$ nucleons are put in $L$
sites by Monte Carlo sampling using the canonical Metropolis
algorithm \cite{Mull97,Metr53}. As pointed
out in Ref. \cite{Carmona,Shida}, however, one has to be careful treating
the process of Metropolis sampling in order to satisfy  the
detailed balance principle and therefore warrant the correct
equilibrium distribution in the final state.
Speaking in detail, in this work, first an initial configuration
with $N + Z$ nucleons is established. Second, for each event,
sufficient number of "spin"-exchange steps is tested, eg. 20000
steps to let the system generate states with a probability
proportional to the Boltzmann probability distribution with
Metropolis algorithm. In each "spin"-exchange step,  a random
trial change on the basis of the previous configuration is made.
For instance, we choose a nucleon at random and attempt to exchange it
 with one of its neighboring nucleons or vacancies
at random regardless of the sign of its "spin" (Kawasaki-like
spin-exchange dynamics \cite{Kawasaki}), then compute the change
$\Delta E$ in the energy of the system due to the trial change.
If $\Delta E$ is less than or equal to zero, accept the new
configuration and repeat the next "spin"-exchange step. If
$\Delta E$ is positive, compute the "transition probability"
$W$ = $e^{-\Delta E/T}$ and  compare it with  a random number
$r$ in the interval $[0,1]$. If $r$ $\leq$ $W$, accept the new
configuration; otherwise retain the previous configuration.
20000 "spin"-exchange steps are performed to assure we get
 the equilibrium state. Third, once the nucleons have been
placed stably on the cubic lattice after 20000 "spin"-exchange
steps for each event, their momenta are generated by Monte Carlo
sampling of Maxwell Boltzmann distribution. Thus various
observables based on phase space
can be calculated in a straightforward fashion for each event.
One important point of  such Monte-Carlo Metropolis
computations is that the above "spin"-exchange approach
between the nearest neighbors, independently of their
"spin", is evidenced to be satisfied by the detailed balance
condition as noted in Ref. \cite{Carmona,Shida}. In other
words,  this sampling method will guarantee that the generated
 microscopic states  form an equilibrium canonical ensemble.

Once this is done the LGM immediately gives the cluster
distribution by using the rule that two nucleons are part
of the same cluster if their relative kinetic energy is
insufficient to overcome the attractive bond \cite{Jpan95}:
\begin{equation}
P_r^2/2\mu - \epsilon_{s_i s_j}s_i s_j < 0.
\end{equation}
This method has been proved to be similar to the so-called
Coniglio-Klein's prescription  in the condensed
matter physics \cite{Coni80} and was shown to be
valid in LGM \cite{Jpan95,Jpan96,Camp97,Gulm98,Ma99}.

\subsection{Isospin dependent classical molecular dynamics
model}

Since the lattice gas model is a model of the nearest neighboring
interaction, a long range Coulomb force is not amenable to lattice
gas type calculation.  Pan and Das Gupta provided a prescription,
based on simple physical reasoning, to decide if two nucleons
occupying neighboring sites, form part of the same cluster or not
\cite{Jpan96}. They first try to map the lattice gas model
calculation to a molecular dynamics type prediction, both first
done without any Coulomb interaction. If the calculations match
quiet faithfully then they can study the effects of Coulomb
interaction by adding that to the molecular dynamics calculation.
Here we adopt the same prescription to use the molecular dynamics
and therefore investigate Coulomb effect. The results and
conclusions can now be compared and checked between the LGM and
the CMD. Obviously, here we do not do any $ab\ initio$ molecular
dynamics calculation but only use it for simulating the nuclear
disassembly starting from a thermally equilibrated source which
has been produced by the above I-LGM: i.e.  the nucleons have been
initialized at their lattice sites with the Metropolis sampling
and have their initial momenta with Maxewell Boltzmann sampling.
From this starting  point we switch the calculation to a classical
molecular dynamics evolution under the influence of a chosen
force. Noting that in this case $\rho_f$ is, strictly speaking,
not a "freeze-out" density for molecular dynamics calculation but
merely defines the starting point for time evolution. However
since classical evolution of a many particle system is entirely
deterministic, the initialization does have in it all the
information of the asympototic cluster distribution, we will
continue to call $\rho_f$ the "freeze-out" density.

The form of the force in the CMD is chosen to be
also isospin dependent in order to compare with the results of I-LGM.
The potential for unlike nucleons is expressed as
\begin{widetext}
\begin{eqnarray}
 v_{\rm n p}(r) (r/r0<a)\ &=&\ C\left[B(r_0/r)^p-(r_0/r)^q\right]\nonumber
    exp({\frac{1}{(r/r_0)-a}}), \\
v_{\rm  n p}(r) (r/r_0>a)\ &=&\ 0.
\label{pot}
\end{eqnarray}
\end{widetext}
where $r_0 = 1.842 fm$ is the distance between the
centers of two adjacent cubes. The parameters of the potentials
are $p$ = 2, $q$ = 1, $a$ = 1.3, $B$ = 0.924, and $C$ = 1966 MeV.
With these parameters the potential is minimum at $r_0$ with the
value -5.33 MeV, is zero when the nucleons are more than 1.3$r_0$
apart and becomes strongly repulsive when $r$ is significantly
less than $r_0$. We now turn to the nuclear potential between like
nucleons. Although we take $\epsilon_{pp}$ = $\epsilon_{nn}$ = 0
in I-LGM, the fact that we don't put two like nucleons in the same
cube would suggest that there is short range repulsion between
them. We have taken the nuclear force between two like nucleons to
be the same expressions as above plus 5.33 MeV up to $r$ = 1.842
fm and zero afterwards:
\begin{eqnarray}
v_{\rm p p}(r) (  r < r0 )\ &=&\  v_{\rm n p}(r)- v
_{\rm n p}(r_0)\nonumber , \\
v_{\rm p p}(r) ( r > r0 )\ &=&\ 0.
\end{eqnarray}
The system evolves with the above potential. The evolution equations
for each nucleon are, as usual, given by
\begin{eqnarray}
 \partial \vec{p_i}/\partial
t\ &=&\ - \Sigma_{j\neq i} \bigtriangledown _i v(r_{ij})\nonumber , \\
\partial \vec{r_i}/\partial t\ &=&\ \vec{p_i}/m.
\end{eqnarray}
Numerically, the particles are propagated in the phase space by
integrating Newton's equations of motion through a "leap-frog"
algorithm. At asymptotic times, for instance, the original blob of
matter has expanded to 64 times its volume in the initialization,
the clusters are easily recognized: nucleons which stay together
after arbitrarily long time are part of the same cluster. The
observables based on cluster distribution in both models can now
be compared. While they can be also compared by switching on/off
Coulomb interaction within the molecular dynamics.

\section{Results and Discussions}

We choose a medium-size nucleus $^{129}$Xe as an example.  The input
parameters are temperature  $T$ and "freeze-out" density $\rho_f$
in the model calculations. In this work $T$ is mostly limited to
the range of 3 to 7 MeV and the "freeze-out" density $\rho_f$ is
chosen to be the range of 0.18$\rho_0$ to 0.60$\rho_0$.
In most cases, $\rho_f$ is chosen to be 0.38 $\rho_0$,  since the
experimental  data can be  best fitted by a $\rho_f$ between
0.3$\rho_0$ and 0.4$\rho_0$ in previous LGM calculations
\cite{Jpan95,Beau96}, which corresponds to  $7 \times 7 \times 7$
cubic lattice for Xe. There is also good
support from experiment that the value of $\rho_f$ is significantly
below 0.5$\rho_0$ \cite{Agos96}. In addition,  0.18$\rho_0$,
corresponding to $9 \times 9 \times 9$ cubic lattice and
0.60$\rho_0$, corresponding to $6 \times 6 \times 6$ cubic
lattice of "freeze-out" densities for $^{129}$Xe are also taken
to compare and check the results with different "freeze-out" density.
1000 events are simulated for each combination of $T$ and $\rho_f$
which ensures good statistics for results.

\subsection{I-LGM in different "freeze-out" density}

Fig.1 shows that the mean multiplicities of emitted neutrons,
protons,  charged particles and intermediate mass fragments and
the  mean mass for the largest fragment evolve with temperature at
different  "freeze-out" densities in the I-LGM calculation. At a
fixed "freeze-out" density, average neutron multiplicity ($N_n$),
proton multiplicity ($N_p$), charged particle multiplicity
($N_{cp}$) and the largest fragment mass ($A_{max}$) display
monotonously increasing or decreasing with  temperature as
expectedly. But the multiplicity ($N_{imf}$) of $IMF$  shows a
rise and fall with temperature \cite{Ogil91,Tsang93,Mayg95},  when
the system probably crosses the phase transition boundary. With
the decreasing of the "freeze-out" density, $N_p$, $N_n$, $N_{cp}$
and $A_{max}$  decreases since larger space separation among
nucleons at smaller "freeze-out" density makes the clusters less
bound and therefore decreases the sizes of free clusters and then
increases the cluster multiplicities.  The situation of $N_{imf}$
is slightly complicate, i.e. it increases with the decreasing of
"freeze-out" density in lower temperature branch but it is
contrary in high temperature branch.
\begin{center}
\begin{figure}
\includegraphics[scale=0.40]{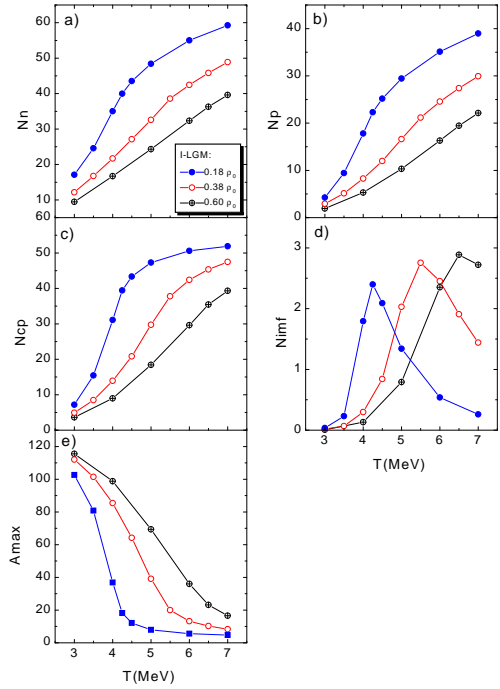}
\caption{\footnotesize Average multiplicity of the emitted
neutrons $N_n$ (a), protons $N_p$  (b), charged particles $N_{cp}$
(c) and intermediate mass fragment $N_{imf}$ (d), average mass of
the largest fragment $A_{max}$ (e) as a function of temperature in
different "freeze-out" density  in the framework of I-LGM.}
\end{figure}
\end{center}

It seems to be impossible to discover the possibility of  phase
 transition of nuclei if we only see these mean quantities
 as shown above ($N_{imf}$ is an exception).
However, when we focus on their slopes to temperature:  figure 2,
the sharp changes are observed at the nearly same
temperature at each fixed  "freeze-out" density: for instance,
4 MeV at 0.18$\rho_0$, 5 MeV at  0.38$\rho_0$ and
6 MeV at 0.60$\rho_0$.  At such a transition point,
(1) the multiplicities of emitted clusters increase rapidly
 and after that the emission  rate slows down; (2) the decrease of the largest
 fragment size reaches to  a valley  for such a finite system.
 Physically the largest fragment is simply related to the order parameter
  $\rho_l$ - $\rho_g$ (the difference of density in nuclear "liquid" and
  "gas" phases). In the infinite matter, the infinite cluster exists only
  on the "liquid" side of critical point. In a finite matter, the largest
  cluster is present on both sides of phase transition point. In this
  calculation, a valley for the slope of $A_{max}$ to temperature may
  correspond to a sudden disappearance of infinite cluster
  ("bulk liquid") near the phase transition temperature.
It is not occasional producing such waves of the slopes,
it should reflect  the onset of  phase transition there.
This idea can be supported by surveying the other phase transition
observables:  such as  the effective power law parameter,
 the Campi's second moment
from the mass or charge  distribution of fragment  and the information
entropy $H$ of event multiplicity distribution \cite{Ma99,Maprl}.
Figure 3 depicts these results.
The minimum of  $\tau$  and the maxima of $S_2$ and $H$ appear around
respective  phase transition temperatures at different values of
"freeze-out" density, $i.e.$   about 4 - 4.25 MeV at 0.18$\rho_0$,
5 - 5.5 MeV at 0.38$\rho_0$  and 6 - 6.5 MeV at  0.60$\rho_0$.
These  temperatures are consistent with the ones  extracted from
the choppy position of the above slopes. It  indicates that the
above slopes (emission rate) can be taken as a probe of a liquid gas
phase transition of  nuclei.
\begin{center}
\begin{figure}
\includegraphics[scale=0.40]{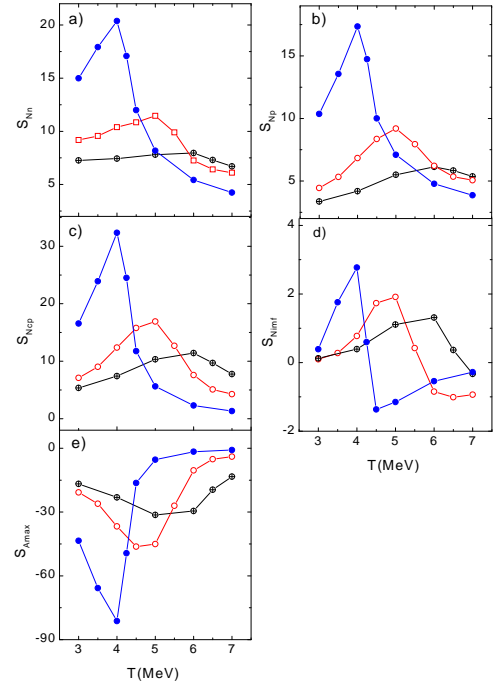}
\caption{\footnotesize Same as figure 1, but for their slopes with
temperature.}
\end{figure}
\end{center}

Furthermore, the largest fluctuations of cluster multiplicities
are found around the phase transition point in the same calculation.
Figure 4 illustrates that RMS width ($\sigma$) of  the multiplicity
distributions of neutrons, protons, charged particles and
intermediate mass fragments, and the distribution of the largest
fragment masses. These variances generally show peaks at the same
phase transition temperatures as the ones extracted from the above
observables for each fixed "freeze-out" density. Noting that the
fluctuation of $A_{max}$ is related to the compressibility of the system.
These features are  also consistent to the one of phase
transition behaviors, i.e.
there exists the largest  fluctuation at the phase transition point
 \cite{Chas96}. This fluctuation represents internal
feature of the disassembling system not numerical fluctuation.
\begin{center}
\begin{figure}
\includegraphics[scale=0.40]{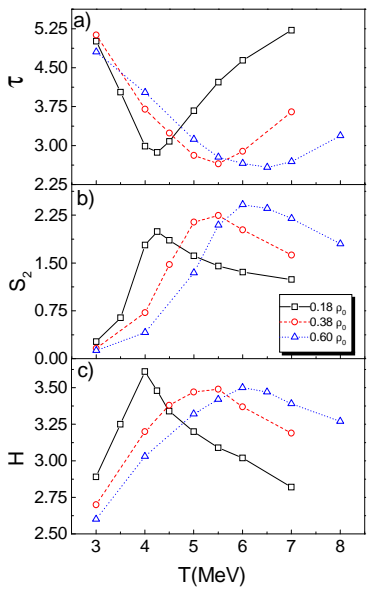}
\caption{\footnotesize Phase transition observables: effective
power law parameter $\tau$ (a), the second moment $S_2$ (b) of
fragment mass distribution and the information entropy $H$ (c) as
a function of temperature in different "freeze-out" density in the
framework of I-LGM.}
\end{figure}
\end{center}

We would like to point that
the "freeze-out" density dependent phase transition temperature,
extracting from $\tau$ and $S_2$,  was also
observed in a previous work \cite{Jpan95}, but this $\rho_f$ dependent
phenomenon will vanish when excitation energy is used as a variable
 \cite{Maepja}. In other words, the excitation energy is perhaps a
 good correspondence of the critical temperature because of  only one
critical point and hence only one critical temperature for a system.
\begin{center}
\begin{figure}
\includegraphics[scale=0.40]{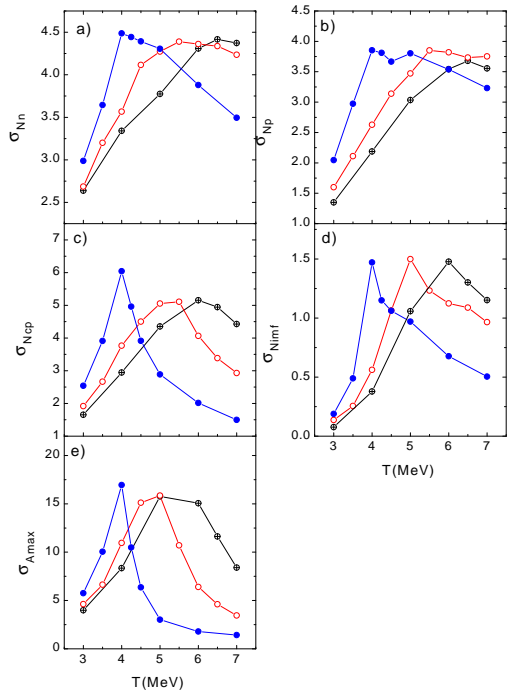}
\caption{\footnotesize Temperature dependent RMS widths ($\sigma$)
of the distributions of $N_n$ (a), $N_p$ (b), $N_{cp}$ (c),
$N_{imf}$ (d) and $A_{max}$ (e)  in different "freeze-out" density
in the framework of I-LGM. The symbols are the same as figure 1. }
\end{figure}
\end{center}

\subsection{Comparison of I-LGM to I-CMD in the same "freeze-out" density}

Considering the absence of long range Coulomb force in the LGM,
we will adopt the CMD to investigate the Coulomb effect and check the
features of cluster emission and its relation to the phase transition
behavior. The same "freeze-out" density of 0.38$\rho_0$ is used to make
a comparison for the results  of I-CMD and I-LGM. Figure 5
shows that $N_p$, $N_n$, $N_{cp}$, $N_{imf}$
and $A_{max}$ changes with temperature within both models.
The multiplicities of
clusters and the largest fragment mass are close each other
between I-LGM and I-CMD except for
the multiplicities of neutrons and protons, illustrating that the I-LGM
is, in general, a good tool to describe the
fragmentation if Coulomb interaction can be ignored.
When Coulomb interaction is switching on, $N_p$, $N_{cp}$ and
$N_{imf}$  increase due to the repulsive role among protons and while
$A_{max}$  decreases. Meanwhile, $N_n$ does not change because of no
Coulomb interaction. $N_{imf}$ also shows a
rise and fall with temperature in the I-CMD cases. Coulomb force makes
the turning temperature of $N_{imf}$ smaller due to the long range repulsion.
\begin{center}
\begin{figure}
\includegraphics[scale=0.40]{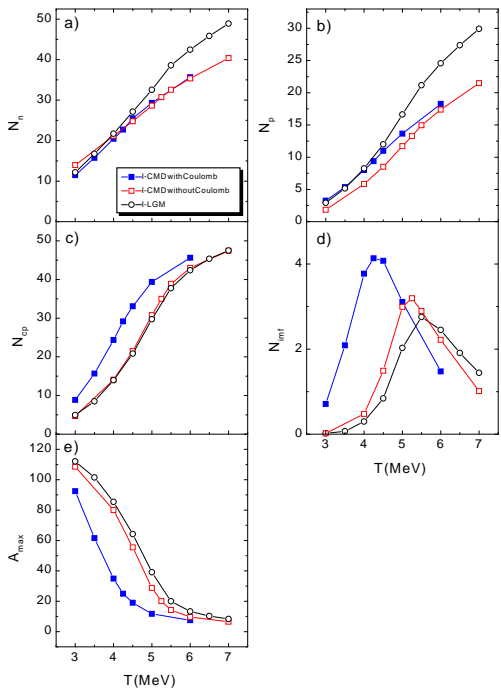}
\caption{\footnotesize  Same as figure 1, but for the comparison
between  different calculations: I-LGM (open circles), I-CMD
without Coulomb interaction (open squares) and I-CMD with Coulomb
force (solid squares). The "freeze-out" density of system is
0.38$\rho_0$. }
\end{figure}
\end{center}

The slopes of multiplicities of emitted clusters and of mean mass
of the largest fragment have been plotted as a function of
temperature in  case of I-CMD in figure 6. The definite peaks of
slopes are found as in the I-LGM case. The corresponding
temperature at peaks locates about 4 MeV in the I-CMD case with
Coulomb interaction, and about 5 MeV in the   I-CMD case without
Coulomb interaction. This turning temperature  should also reflect
the onset of  phase transition there. If we investigate $\tau$,
$S_2$ and $H$: Figure 7, we will find that there are a  minimum of
$\tau$ and the maxima of $S_2$ and $H$ around phase transition
temperatures, $i.e. $  about 4 - 4.25 Mev for I-CMD with Coulomb,
around 5 MeV for I-CMD without Coulomb and  around 5.5 MeV for
I-LGM, respectively. Obviously, these phase transition
temperatures are consistent with the ones extracted from the
choppy position of the slopes of figure 6.
\begin{center}
\begin{figure}
\includegraphics[scale=0.40]{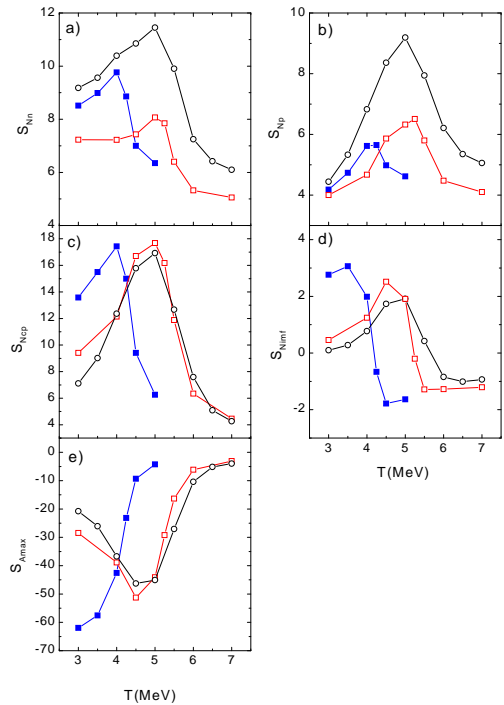}
\caption{\footnotesize Same as figure 2, but for the comparison
between different calculations: I-LGM (open circles), I-CMD
without Coulomb interaction (open squares) and I-CMD with Coulomb
force (solid squares). The "freeze-out" density of system is
0.38$\rho_0$.}
\end{figure}
\end{center}

Finally the RMS widths of the multiplicity distributions of the
clusters and of the largest fragment mass are checked in the I-CMD
case in figure 8.  The widths of the multiplicity distributions of
neutrons and protons tend to be saturate at higher temperature,
and while the ones for $CP$, $IMF$ and $A_{max}$ demonstrate
peaks at a certain fixed temperature, i.e. around 4 MeV for the
case I-CMD with Coulomb, 5 MeV for the case of I-CMD without
the Coulomb, which is similar to the I-LGM case. These turning
temperatures are also consistent to the phase transition temperature
as shown in figure 6 and 7 in the I-CMD cases. Again, the largest
fluctuation simultaneously appears in the  point of phase transition.
\begin{center}
\begin{figure}
\includegraphics[scale=0.40]{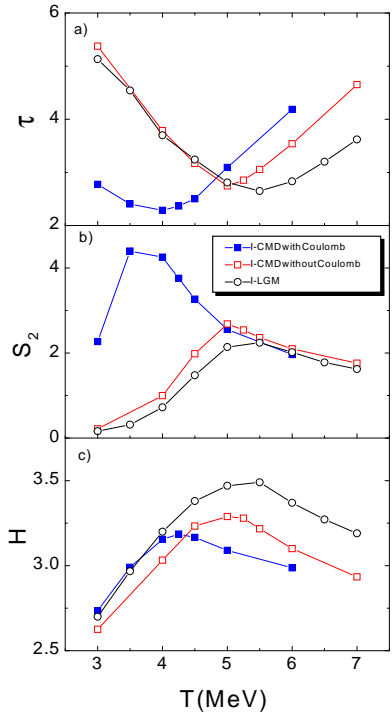}
\caption{\footnotesize Same as figure 3, but for the comparison
between different calculations: I-LGM (open circles), I-CMD
without Coulomb interaction (open squares) and I-CMD with Coulomb
force (solid squares). The "freeze-out" density of system is
0.38$\rho_0$.}
\end{figure}
\end{center}

Overall, the phase transition  temperature seems to rely
 on some ingredients:
such as  the "freeze-out" density (or pressure), the model and its
interaction potential, but the rule that emission rate and
 fluctuation of cluster multiplicity as a probe of phase transition
is not changed.
\begin{center}
\begin{figure}
\includegraphics[scale=0.40]{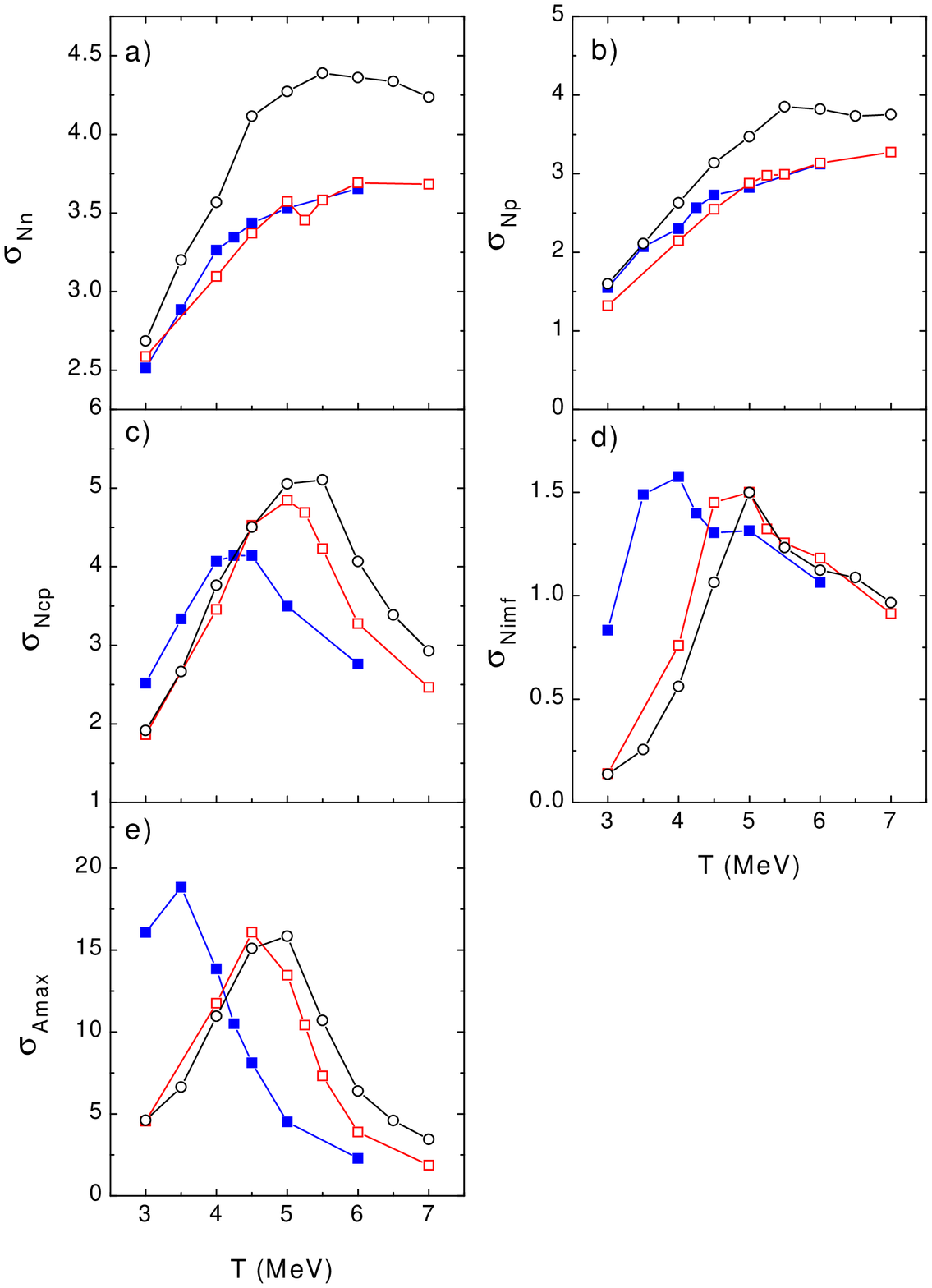}
\caption{\footnotesize Same as figure 4, but for the comparison
between different calculations: I-LGM (open circles), I-CMD
without Coulomb interaction (open squares) and I-CMD with Coulomb
force (solid squares). The "freeze-out" density of system is
0.38$\rho_0$.}
\end{figure}
\end{center}
\section{Conclusions}

In conclusion, features of the emissions of light particles, charged particles,
intermediate mass fragments and the largest fragments are investigated
for a medium size nucleus $^{129}Xe$ in the frameworks of isospin
dependent lattice gas model and classical molecular dynamics model.
$N_n$, $N_p$, $N_{cp}$ and $A_{max}$  show monotonously
increasing or decreasing but $N_{imf}$ shows a rise and fall with
temperature. Slopes of these observables go through  extrema
at the same  temperature  where the largest  fluctuation of cluster
multiplicity distributions is also observed. This  temperature
is  consistent to the phase transition temperature
extracted from the extreme values of $\tau$, $S_2$ and $H$.
It gives an indication that the cluster emission rate
can be taken as a probe of the phase transition of nuclei and while
the largest  fluctuation is simultaneously accompanied when the
onset of phase transition occurs.
In addition, the systematical comparison of I-LGM and I-CMD shows that
LGM is a rather good tool to study the nuclear disassembly when the system
is not large where the Coulomb interaction can be ignored.
In light of this work, we think that the experimental study about
the cluster emission is rather meaningful, especially to measure the
excitation function of the multiplicities, their slopes and  variances,
from which some signals for phase transition  could be found.

\vspace{.3cm}

\centerline{ACKNOWLEDGEMENT}
\vspace{.3cm}

It's my pleasure to thank Prof. S. Das Gupta and Dr. J. Pan for
kindly providing  the original codes and Prof. B. Tamain for
helps. This work was partly supported  by the NSFC for
Distinguished Young Scholar under Grant No. 19725521, the NSFC
under Grant No. 19705012, the Presidential Foundation of Chinese
Academy of Sciences, the Science and Technology Development
Foundation of Shanghai under Grant No. 97QA14038.

{}

\end{document}